\documentclass[aps,prl,showpacs,twocolumn,amsmath,amssymb,superscriptaddress]{revtex4-2}

\usepackage{tabularx}
\usepackage{bm}
\usepackage{graphicx}
\usepackage{comment}

\usepackage{hyperref}
\hypersetup{colorlinks=true,urlcolor= blue,citecolor=blue,linkcolor= blue,bookmarks=true,bookmarksopen=false}

\usepackage{color}

\usepackage{amsmath,mathtools}
\usepackage{multirow}
\usepackage{dcolumn}
\usepackage{amssymb,amscd,xypic,bm,wasysym}
\usepackage{float}
\usepackage{cleveref}
\usepackage[caption=false,position=top,captionskip=0pt,farskip=0pt]{subfig}
\usepackage{soul}

\begin{document}
	
\title{Switchable chiral antiferromagnetism through nonlinear magnetic susceptibility}
	
\author{Hua Chen}
\email{huachen@colostate.edu}
\affiliation{Department of Physics, Colorado State University, Fort Collins, CO 80523, USA}
\affiliation{School of Advanced Materials Discovery, Colorado State University, Fort Collins, CO 80523, USA}
\author{Philipp Gegenwart}
\affiliation{Experimentalphysik VI, Center for Electronic Correlations and Magnetism, University of Augsburg, 86159 Augsburg, Germany}
%\author{Kan Zhao}
%\affiliation{School of Physics, Beihang University, Beijing 100191, China}
%\affiliation{Experimentalphysik VI, Center for Electronic Correlations and Magnetism, University of Augsburg, 86159 Augsburg, Germany}

\begin{abstract}
Antiferromagnets (AFM) have attracted considerable attention in recent years because a number of nontrivial, technologically relevant properties associated with time-reversal symmetry (TRS) breaking are discovered in many materials. However, time-reversal (TR) partners of AFM states are generally challenging to be selected deterministically except in a few cases with nonzero net magnetization. Recently, it has been shown that the TR partners of the kagome spin ice ground state in HoAgGe with strictly zero net magnetization can be selected through a nonlinear magnetic susceptibility $\chi^{(1)}$. In this work, we generalize this scenario to a broad class of chiral AFM with zero net magnetization but field-switchable ground states through $\chi^{(1)}$. After a general discussion of $\chi^{(1)}$ and its symmetry constraints, we first present a first-principles formalism for calculating it including both mean-field and self-consistent corrections, and then apply the approach to the noncollinear AFM family Mn$_3X$N ($X=$ Ni, Ag, Ga, Zn, Sn) in the $\Gamma_{5g}$ phase. An intuitive picture of the origin of $\chi^{(1)}$ in similar noncollinear AFM is illustrated using 3-sublattice toy model. The model also predicts a nontrivial temperature dependence of $\chi^{(1)}$ due to competitions between the longitudinal and transverse single-spin susceptibilities. Our work shows that nonlinear susceptibility can serve as a general protocol for accessing and switching TR partners in fully compensated AFM.

\end{abstract}

\maketitle

\textit{Introduction.---}Interests in antiferromagnets (AFM) are surging due to many discoveries in recent years that demonstrate their time-reversal symmetry (TRS) breaking has nontrivial consequences in transport, magnetic, and optical properties \cite{Solovyev1997, Tomizawa2009, Ohgushi2000, Shindou2001, Chen2014, Kuebler2014, Nakatsuji2015, Nayak2016, Zelezny2014, Wadley2016, Zhou2019, Gurung2019, Boldrin2019, Zhao2019, Liu2018, Smejkal2020, Chen2020, Chen2022}. Examples include the anomalous Hall effect \cite{Chen2014, Kuebler2014, Nakatsuji2015, Nayak2016}, spin-polarized spin currents \cite{Zelezny2017} and the closely related magnetic spin Hall effect \cite{Kimata2019,Kondou2021}, non-relativistic spin splitting in electronic and magnonic bands \cite{Smejkal2022a, Smejkal2022, Liu2022, Guo2025, Shim2025, Tamang2025, Song2025, Jungwirth2026}, tunneling magnetoresistance \cite{Chen2023,Qin2023}, piezomagnetic response \cite{Ikhlas2022,ZunigaCespedes2023}, etc. These developments have recently culminated in the bourgeoning field of altermagnetism \cite{Smejkal2022a, Smejkal2022, Liu2022, Guo2025, Shim2025, Tamang2025, Song2025, Jungwirth2026}. 

However, most proposals of unique phenomena of AFM require or at least benefit from having a single AFM domain. Except in a few cases where the net magnetization is not forbidden by symmetry, such as Mn$_3X$ ($X$ = Ir, Pt, Sn, Ge, etc.) and other weak ferromagnets, whose time-reversal (TR) partners of the AFM ground states can be energetically biased using external magnetic fields, there lacks a general, viable strategy for achieving a single AFM domain in a finite sample. Another promising approach from spintronics is to employ a current-induced staggered spin-orbit torque \cite{Zelezny2014, Wadley2016}, which, however, is limited to metals and those with local inversion symmetry breaking about the positions of magnetic ions.

In a recent work involving the present authors \cite{Zhao2026}, it has been shown that the $\sqrt{3}\times \sqrt{3}$ noncollinear ground state of the kagome spin ice compound HoAgGe \cite{Zhao2020}, which has net magnetization $M=0$, can be switched between its two TR partners through a nonlinear magnetic susceptibility $\chi^{(1)} = \frac{\partial^2 M}{\partial B^2}$ when the magnetic field is applied along the in-plane $b$ axis, parallel to a Ho magnetic moment in the ground state. Specifically, sweeping the magnetic field $B_b$ creates a hysteresis of the $\chi^{(1)}(B)$ curve, similar to the $M(B)$ hysteresis in a ferromagnet. A microscopic mechanism of $\chi^{(1)}$ unique to kagome spin ice systems is given in \cite{Zhao2026}, which is that the ice-rule-compliant 1-spin-flip excitations from the two TR partners of the ground state become non-degenerate under finite $B_b$. 

\begin{figure}[h]
    \centering
         \includegraphics[width=0.3\textwidth]{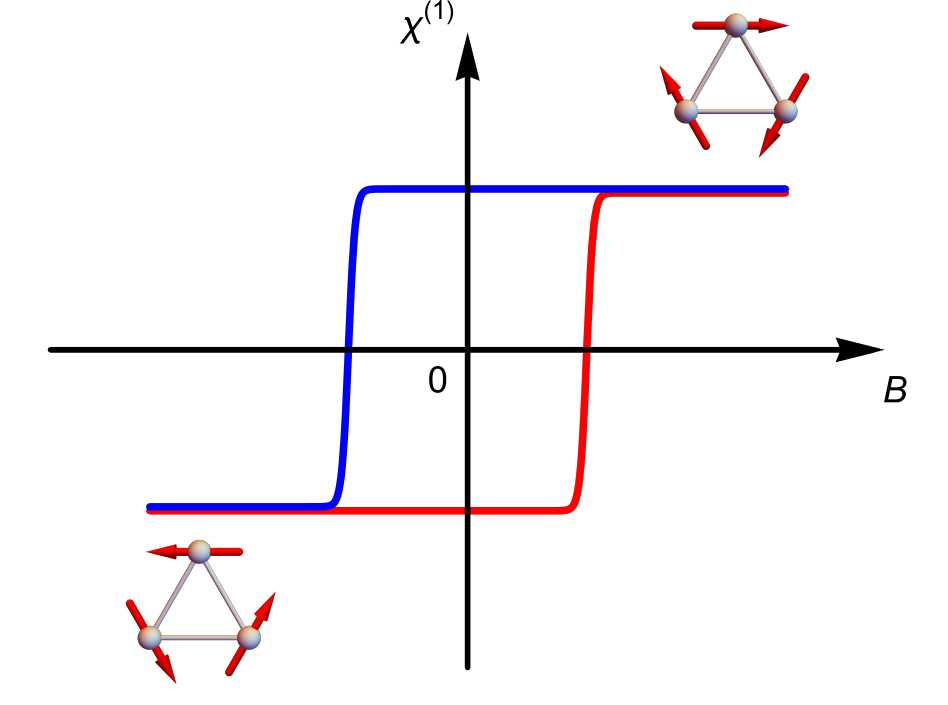}   
    \caption{Schematic of field-switchable AFM through the hysteretic nonlinear susceptibility $\chi^{(1)}(B)$. Insets show the two TR partners of the $\Gamma_{5g}$ phase of Mn$_3X$N. The applied $\mathbf B$ is along or opposite to one of the spins.}
    \label{fig:schematic}
\end{figure}

Nonlinear magnetic susceptibilities and the resulting field-induced hysteresis have only seen scattered discussions in literature, dating back to 1970s, and are mostly limited to noncollinear Ising spin systems \cite{Giordano1980,Gorodetsky1967,Kharchenko1995,Gregg1990,Wolf1990,Alben1975,Foglio1977,Mukamel1977,Blume1974,Dillon1974,Fujita2015}. In particular, a first-principles framework for the nonlinear susceptibility applicable to general AFM materials is lacking. Understanding and quantifying nonlinear susceptibility in AFM will be instrumental in promoting AFM research since the otherwise degenerate TR partner states can be deterministically selected and characterized using standard magnetic-field techniques developed for ferromagnets (Fig.~\ref{fig:schematic}). 

In this Letter, we provide a comprehensive first-principles theory of nonlinear susceptibility in general AFM materials. Following a general introduction to nonlinear susceptibility and its symmetry constraints, we first give analytic formulas of $\chi^{(1)}$ in mean-field electronic systems and strategy for obtaining the self-consistent-field (scf) correction. Applying the formalism to the family of Mn$_3X$N ($X$=Ni, Ag, Sn, Zn, Ga) in the $\Gamma_{5g}$ phase \cite{Fruchart1978,phdthesis}, we find that the scf corrections are dominant over the bare mean-field $\chi^{(1)}$. We therefore focus on the scf part of $\chi^{(1)}$, which bears an intuitive physical explanation in similar noncollinear AFM systems that can be described using spin models. Finally, using a mean-field treatment, we predict an intriguing sign change of $\chi^{(1)}$ versus temperature, due to the competition between transverse and longitudinal susceptibilities of local spins. Finally, we discuss further implications of our work.

\textit{General introduction to nonlinear susceptibility.---}A magnetically ordered state $S$ and its time-reversal partner $S'$ are always degenerate in the absence of external perturbations. Under a uniform magnetic field, their degeneracy can be lifted provided that the free energy or thermodynamic potential of the two states satisfy
\begin{equation}\label{eq:switchFneq}
    \Omega_{S}(\mathbf B) \neq \Omega_{S'}(\mathbf B) = \Omega_{S}(-\mathbf B)
\end{equation}
The inequality therefore means that the series expansion of $\Omega_S$ or $\Omega_{S'}$ with respect to $\mathbf B$ must have odd-power terms. Up to the third order:
\begin{eqnarray}\label{eq:Fexpansion}
\Omega_S \approx \Omega_0 - M_{i} B_{i} - \frac{1}{2}\chi^{(0)}_{ij} B_{i }B_{j} - \frac{1}{6} \chi^{(1)}_{ijk} B_{i }B_{j} B_{k}
\end{eqnarray}
where $\mathbf M$ is the net magnetization, $\chi^{(0)}$ is the usual (linear) magnetic susceptibility at zero magnetic field, and $\chi^{(1)}$ is the lowest order nonlinear susceptibility. For AFM with strictly vanishing $\mathbf M$, a nonzero $\chi^{(1)}$ or more generally, any odd-order nonlinear susceptibility $\chi^{(2n+1)}$ ($n\geq 0$)  can still allow $S$ and $S'$ to be switched into each other by $\mathbf B$.

Since $\mathbf B$ is a TR-odd pseudovector, $\chi^{(2n+1)}$ vanish identically if $S$ has either TR ($\mathcal{T}$) symmetry or that combined with spatial inversion ($\mathcal{I}$) symmetry. Conversely, AFM states without $\mathcal{T}$ (combined with lattice translation) and $\mathcal{TI}$ symmetry are in general possible to be switched by a uniform magnetic field along certain directions for which $\chi^{(2n+1)}$ is nonzero. It is, of course, possible that the given $\mathbf B$ can induce other metamagnetic transitions, such as spin flip/flop transitions for collinear AFM, and more complicated situations for noncollinear AFM. Such cases are more system-dependent and will only be discussed briefly at the end.

In this work, we focus on the lowest order nonlinear susceptibility $\chi^{(1)}$, with $\mathbf B$ applied along a high-symmetry direction, chosen as $z$. We also consider the contribution to $\chi^{(1)}$ in the spin sector only, since typical antiferromagnetism is driven by spin order. Nonetheless, the discussion below can be straightforwardly generalized to that for magnetic fields coupled to the orbital degrees of freedom, for which a nonlinear version of the orbital-spin susceptibility \cite{Chen2020} is relevant. 

$\chi_{zzz}^{(1)}$ can be further constrained by other magnetic or spin space group symmetries. For example, we showed in \cite{Zhao2026} that a nonzero $\chi_{zzz}^{(1)}$ additionally requires the absence of $C_{2x,y}, m_{x,y}, C_{2z}\mathcal{T}, m_{z}\mathcal{T}$ magnetic space group symmetries. Moreover, in the absence of spin-orbit coupling (SOC), $\chi^{(1)}$ can still be nonzero in general (see below). For collinear AFM in the zero SOC limit, rotation about the collinear axis $\hat{n}$ by any angle is a symmetry, which forbids all components of $\chi^{(1)}$ except that along $\hat{n}$. This can be undesirable since the field is likely to cause a spin-flop transition. For coplanar but noncollinear AFM with $\hat{n}$ normal to the ordering plane, a $\pi$ rotation about $\hat{n}$ followed by $\mathcal{T}$ is a symmetry. $\chi^{(1)}$ then can only have components perpendicular to $\hat{n}$. 

The above symmetry constraints and their generalizations can be readily used to screen $\chi^{(1)}$-switchable AFM materials based on their magnetic and/or spin space groups \cite{Brinkman1966, Litvin1974, Litvin1977, Liu2022, Chen2024, Jiang2024, Xiao2024, Watanabe2024, Etxebarria2025, Liu2025, Liu2026}, which will be left for a future work. Below we focus on the microscopic mechanisms leading to $\chi^{(1)}$ and try to gain a more quantitative understanding of it in some topical AFM materials.

\textit{Electronic contribution to $\chi^{(1)}$.---}We next give a microscopic theory of $\chi^{(1)}$ for an electronic system. In the spirit of spin density functional theory \cite{MacDonald1979, Crepieux2001}, we first calculate the independent-electron or mean-field contribution to $\chi^{(1)}$, and then discuss the self-consistent-field (scf) correction.

For the mean-field part, we use a similar approach to that in \cite{Chen2020} and directly expand the free energy in powers of $\mathbf B$. Specializing to $\chi^{(1)}_{zzz}$, we obtain the following Fermi-surface and Fermi-sea contributions to $\chi^{(1)}$ \cite{supp, Altland2023}:
\begin{widetext}
\begin{eqnarray}\label{eq:chi1surfseadiag}
    \chi^{(1),\rm surf}
    &=& -{\mu_B^3}\int_{\rm BZ} \frac{d^3\mathbf k}{(2\pi)^3}\left(\sum_a f_a''(s_{aa})^3 + 3 \sum_{a\neq b}  \frac{s_{aa}  {f'_a} - s_{bb}  {f'_b}}{\epsilon_{ab}} |s_{ab}|^2\right)\\\nonumber
    \chi^{(1),\rm sea} &=& -{2\mu_B^3}\int_{\rm BZ} \frac{d^3\mathbf k}{(2\pi)^3}\left[- 3 \sum_{a\neq b}  s_{aa} |s_{ab}|^2 \frac{f_{ab}}{\epsilon_{ab}^2} -\sum_{a\neq b\neq c\neq a}  {\rm Re} (s_{ab} s_{bc} s_{ca})  \frac{f_a \epsilon_{bc} + f_b \epsilon_{ca} + f_c \epsilon_{ab}}{\epsilon_{ab}\epsilon_{bc}\epsilon_{ca}}  \right]
\end{eqnarray}
\end{widetext}
where $a,b,c$ label Bloch bands, $f_a = f(\epsilon_{a\mathbf k})$ is the Fermi-Dirac distribution function for band $a$, $f'_a = \partial f(\epsilon_{a\mathbf k})/\partial \epsilon_{a\mathbf k}$, $f_{ab} = f_a - f_b$, $\epsilon_{ab} \equiv \epsilon_{a\mathbf k} - \epsilon_{b\mathbf k}$, $s_{ab} = \langle u_{a\mathbf k}| \hat{s}^z|u_{b\mathbf k}\rangle$, with $\hat{\mathbf s}$ the Pauli matrix vector.

Eq.~\eqref{eq:chi1surfseadiag} can be better understood by comparing it to the linear susceptibility:
\begin{eqnarray}\label{eq:chi0}
    \chi^{(0)}_{\alpha\beta} = -{\mu_B^2}\int_{\rm BZ} \frac{d^3\mathbf k}{(2\pi)^3}\sum_a  \left(\frac{\partial f_a}{\partial \Delta^\beta} s^\alpha_{aa} + f_a \partial_{\Delta^\beta} s^\alpha_{aa} \right)
\end{eqnarray}
where $\boldsymbol \Delta$ is a field conjugate to $\mathbf s$. The two terms are respectively the Pauli susceptibility $\chi_{\rm P}$ (field-induced spin-dependent energy shift) and the Van Vleck susceptibility $\chi_{\rm V}$ (field-induced perturbation of eigenstates). The first term in $\chi^{(1),\rm surf}$ comes from $\chi_{\rm P}$ at finite fields or equivalently the $O(B^2)$ energy shift that is spin dependent. The second term in $\chi^{(1),\rm surf}$ has two parts from $\chi_{\rm P}$ and $\chi_{\rm V}$, respectively, but both can be understood as the $\chi_{\rm V}$ of the field-induced net spin at the Fermi surface. For $\chi^{(1),\rm sea}$, both terms can be understood as due to higher-order perturbations of the eigenstates similar to that in $\chi_{\rm V}$.

For collinear magnets without SOC, Eq.~\eqref{eq:chi1surfseadiag} reduces to
\begin{eqnarray}
    \chi^{(1)} = \chi^{(1),\rm surf} = -{\mu_B^3} \partial_\mu \left(D^{\uparrow} -  D^{\downarrow}\right)
\end{eqnarray}
Namely, $\chi^{(1)} $ only has a Fermi surface contribution originating from the chemical-potential dependence of opposite-spin density of states at the Fermi level. This applies to, e.g. rare-earth-cobalt alloys at the compensation point \cite{Buschow1980, CHOE1987, Hansen1989, Finley2016, Xiao2021, Ard2025}, where the net magnetization vanishes, but $\chi^{(1)}$ can still be finite and potentially make the AFM states field-switchable.

Although Eq.~\eqref{eq:chi1surfseadiag} can be used for model \cite{supp} and density functional theory (DFT) calculations, it misses the scf corrections that can be significant in typical AFM materials. Specifically, the full response to the Zeeman field $\mathbf B$ should take the self-consistent dependence of the ordered spin densities on $\mathbf B$ into account. In particular, for AFM whose magnetic properties can be well understood using spin Hamiltonians, the scf contribution, which corresponds to the deformation of the ordered spin configurations due to the finite $\mathbf B$, is expected to be the dominant mechanism for $\chi^{(1)}$.

Here we propose a simple approach to calculating $\chi^{(1)}$ including the scf correction in standard DFT codes. We perform self-consistent calculations for the given system under a few small Zeeman fields symmetric about 0, e.g.,
\begin{eqnarray}
    \mathbf B^+_i = B_i \hat{z} = -\mathbf B^-_i, \quad B_i = \frac{i}{N_B} B_{\rm max}
\end{eqnarray}
and obtain the corresponding net magnetization $\mathbf M(B_i^{\pm})$. The symmetric part of $M_z(B_i)$,
    $M^s_i \equiv \left[ M_z(B_i^+) + M_z(B_i^-)\right]/2$
    satisfies
\begin{eqnarray}\label{eq:chi1regression}
    M^s_i = \frac{1}{2}\chi^{(1)} B_i^2 + \epsilon
\end{eqnarray}
from which $\chi^{(1)}$ can be solved as a standard nonlinear regression problem \cite{supp}. Model calculations show that the fitting approach gives virtually identical results to the analytic formula Eq.~\eqref{eq:chi1surfseadiag}. 

\textit{$\chi^{(1)}$ of Mn$_3X$N.---}We now use the above scheme to calculate $\chi^{(1)}$ of the antiperovskite nitride family Mn$_3X$N, ($X$=Ni, Ag, Sn, Zn, Ga), which have garnered a lot of interest recently due to their noncollinear AFM order, strong magnetostriction/piezomagnetism, and (for Mn$_3$NiN) the anomalous Hall effect in the $\Gamma_{4g}$ phase similar to that in cubic Mn$_3X$ \cite{Fruchart1978, Takenaka2005, Asano2008, Ding2011, Shibayama2011, Takenaka2014, Matsunami2014, Boldrin2018, Shi2018, Zhou2019, Gurung2019, Boldrin2019, Zhao2019, Singh2021, Farhang2026, Chen2026}. These compounds are isostructural in the paramagnetic phase with space group {Pm-3m} (No.~221). They also share the same noncollinear magnetic structure ($\Gamma_{5g}$) with magnetic space group $\text{R-3m.1}$ (No.~166.97) in different low-temperature regions. The magnetic unit cell is identical to the structural one and has three Mn spins aligned with face diagonals of the cubic unit cell, adding up to zero net spin. Symmetry analysis with or without SOC shows that $\chi^{(1)}$ is nonzero for the component parallel to any Mn spin \cite{Aroyo2006, Aroyo2006a, Gallego2019}. 

DFT calculations are performed using Quantum ESPRESSO \cite{Giannozzi2009, Giannozzi2017}, and fully-relativistic ultrasoft PBEsol pseudopotentials contained in PSlibrary \cite{DalCorso2014}. The scf calculations use cutoff values of $80~\rm Ry$ and $800~\rm Ry$ for wavefunctions and densities, respectively. To accurately calculate the net magnetization under finite Zeeman fields, we use scf convergence threashold of at least $5\times 10^{-9}$ Ry, $20\times 20\times 20$ $k$-mesh, and the optimized tetrahedron method \cite{Kawamura2014, Bloechl1994a}.

Table~\ref{tab:Mn3XNchi01} lists the calculated $\chi^{(0)}$ as well as $\chi^{(1)}$ of Mn$_3X$N. Among them, Mn$_3$AgN has the largest value, reaching $\sim -6\times 10^{-3}$ $\mu_B\rm T^{-2}$. Therefore, at relatively moderate fields of 1 T, the two TR partners differ in their net magnetization by $\sim 10^{-2}$ $\mu_B$ per unit cell, which is of similar order of magnitude as that in weak ferromagnets at zero field, such as Mn$_3$Sn, and can lead to deterministic switching especially at elevated temperatures when the energy barrier along the switching path is comparable to thermal energy. We also note that the values in Table~\ref{tab:Mn3XNchi01} may be underestimated compared to experimental values as discussed in Appendix.

\begin{table}[ht]
    \centering
    \renewcommand{\arraystretch}{1.5} % increased row spacing
     \setlength{\tabcolsep}{4pt}
    \begin{tabular}{c | c | c | c | c  }
    \hline\hline 
     \multirow{2}{*}{$X$} & \multicolumn{2}{c|}{$\chi^{(0)}$ ($\mu_B\rm T^{-1}$)} & \multicolumn{2}{c}{$\chi^{(1)}$ ($\mu_B\rm T^{-2}$)}\\ \cline{2-5}
     & nscf ($10^{-4}$) & scf ($10^{-3}$) & nscf ($10^{-9}$) & scf ($10^{-6}$) \\\hline
     Ni & $3.0966(8) $  & $1.740(5) $ & $4.64(2)$ & $1.3(3)$ \\
    Ag & $3.06(1) $ & $6.5(7)$ & $-5.6(4)$ & $-6.4(3)\times 10^{3}$ \\
   Ga & $2.6673(1)$  & $1.247(4)$ & $1.29(2)$ & $0.764(7)$ \\
   Zn & $2.7243(5)$  & $1.531(5)$ & $-2.399(8)$ & $1.00(1)$ \\
   Sn & $2.93(1)$   & $6.26(3)$ &  $6.9(4)$& $-1.2(1)\times 10^{2}$ \\ 
    \hline\hline
    \end{tabular}
    \caption{Linear ($\chi^{(0)}$) and nonlinear ($\chi^{(1)}$) susceptibilities of Mn$_3X$N per unit cell along $[1\bar{1}0]$ (the direction of an Mn spin) in the $\Gamma_{5g}$ phase calculated with and without scf corrections. } 
    \label{tab:Mn3XNchi01}
\end{table}

Another interesting observation from Table~\ref{tab:Mn3XNchi01} is that the scf-corrected $\chi^{(1)}$ are at least two orders of magnitude larger than the bare values. Therefore the scf correction dominates $\chi^{(1)}$ in these materials at zero temperature. Since the scf correction can be understood as deformation of ordered spin configuration, it strongly suggests that the microscopic origin of $\chi^{(1)}$ in Mn$_3X$N and similar noncollinear AFM materials can be captured by spin models at least qualitatively, as we discuss next. 

\textit{$\chi^{(1)}$ from a noncollinear spin model.---}We introduce a minimal 3-sublattice spin model to illustrate the origin of $\chi^{(1)}$ and also gain an initial understanding of its temperature dependence. The model has three Heisenberg spins with nearest-neighbor AFM exchange coupling $J$ forming a $120^\circ$ order, parallel to the $zx$ plane. Each spin in sublattice $a$ is assumed to have $N_c$ nearest neighbors in a different sublattice $b$ ($N_c = 4$ for Mn$_3X$N). A sublattice-dependent easy-axis anisotropy of size $K$ consistent with the $120^\circ$ order is also included. The spin Hamiltonian with an additional external magnetic field $\mathbf B = B\hat{z}$ is
\begin{eqnarray}
    H=J\sum_{\langle ia,jb\rangle}\hat{n}_{ia}\cdot \hat{n}_{jb} -\frac{K}{2}\sum_{ia}(\hat{m}_a \cdot \hat{n}_{ia})^2 - BM\sum_{ia}\hat{n}_{ia}\cdot \hat{z}
\end{eqnarray}
where $\hat{n}_{ia}$ are unit vectors along the local magnetic moment direction on sublattice $a$ in unit cell $i$, $\hat{m}_a$ are the easy axes forming a $120^\circ$ triad, with $\hat{m}_1 =\hat{z}\parallel \mathbf B$.

We first consider the zero temperature case, for which $\chi^{(1)}$ can be calculated by minimizing the ground state energy at finite $B$, analogous to the scf correction in DFT calculations. Due to symmetry, there is only one variational parameter, which is the canting angle $\phi$ away from $\hat{m}_a$ for spins in sublattices $2,3$ towards $\hat{z}$. In the absence of $K$, the ground state energy has only a $B^2$ term which gives the familiar result $\chi^{(0)} = \frac{M^2}{JN_c}$. Therefore, the pure Heisenberg model does not have $\chi^{(1)}$. 

To understand the qualitative change induced by $K$, we treat it as a perturbation ($K\ll B<J$) and calculate its 1st-order correction to $H$ \cite{supp}:
\begin{eqnarray}
    \delta H = -K\cos^2\phi \approx \left(-1 + \frac{x^2}{3} - \frac{x^3}{9}\right)K
\end{eqnarray}
where $x\equiv MB/JN_c$. Therefore
\begin{eqnarray}\label{eq:chi1toy0K}
    \chi^{(1)}\approx \frac{2 K M^3 }{3 (JN_c)^3}
\end{eqnarray}

The reason why $\chi^{(1)}$ is induced by the anisotropy can be understood in the following way. In the absence of $K$, the vanishing $\chi^{(1)}$ means that net magnetization $M(B) = -M(-B)$. Since $M(B)$ is due to the canting of spins in sublattices $2,3$, whose easy axes are not perpendicular to $\hat{z}$, one must have $\phi(B)\neq -\phi(-B)$, which makes anisotropy energy $\propto \cos^2\phi$ different between $\pm B$.

At finite temperatures, $\chi^{(1)}$ has additional contributions from electronic and magnonic excitations. These contributions are further complicated by the fact that the mean-field ground states that they are based on is also perturbed by the external magnetic field in a self-consistent manner, for which phonons may also participate. The bare thermal excitation contributions are generally expected to have non-monotonic temperature dependence, since they must vanish at both $T = 0$ K and at the transition temperature $T_c$. The latter is because $\chi^{(1)}$ must vanish in the paramagnetic state. Therefore, it is possible that the experimentally measured $\chi^{(1)}$ has sizable enhancement at finite temperatures. For example, in HoAgGe, $\chi^{(1)}$ is as large as $0.287~\mu_B\rm T^{-2}$ per formula unit at 6 K but rapidly diminishes to undetectable as temperature approaches 0. 

Here we use the toy model to illustrate another nontrivial feature of the temperature dependence of $\chi^{(1)}$, potentially applicable to a broad class of noncollinear AFM: $\chi^{(1)}$ can change sign between $0$ K and $T_c$. To see this we use an scf approach and write the ordered $\hat{n}$ on the three sublattices as $B$-dependent vectors:
\begin{eqnarray}
    \mathbf n_1  = a \hat{z},\;\mathbf n_{2,3}  = b \hat{z} \pm c \hat{x}
\end{eqnarray}
The mean-fields $a,b,c$ are determined from self-consistent equations of the form
\begin{eqnarray}
   \mathbf n = \frac{\int d\hat{n} \hat{n} e^{ \hat{n}\cdot \mathbf h}}{\int d\hat{n} e^{\hat{n}\cdot \mathbf h}} = \frac{\mathbf h}{h} L(h)
\end{eqnarray}
where $L(h) = \frac{1}{\tanh (h)} - \frac{1}{h}$ is the Langevin function, and $\mathbf h$ is the effective field coupled to the dynamical spin unit vector $\hat{n}$, obtained from the mean-field-decoupled Hamiltonian. $\chi^{(1)}$ is calculated by expanding the net magnetization $(a+2b)M$ to $O(B^2)$. As a sanity check, calculating $\chi^{(0)}$ and $\chi^{(1)}$ at $T\rightarrow 0$ K and $K\ll J$ gives \cite{supp}
\begin{eqnarray}\label{eq:chi0chi1lowT}
    \chi^{(0)}=  \left(1 - \frac{2K}{3JN_c}\right) \frac{M^2}{JN_c},~ \chi^{(1)} = \frac{2K M^3}{3(JN_c)^3 n}
\end{eqnarray}
where $n = |\mathbf n|$ is the ordered spin without $\mathbf B$ on each sublattice. $\chi^{(0)}$ is suppressed by a finite $K$ as expected. More interestingly, $\chi^{(1)}$ increases with temperature from the $T=0$ result Eq.~\eqref{eq:chi1toy0K} as $1/n(T)$. 

\begin{figure}[h]
    \centering
         \subfloat[]{\includegraphics[width=0.3\textwidth]{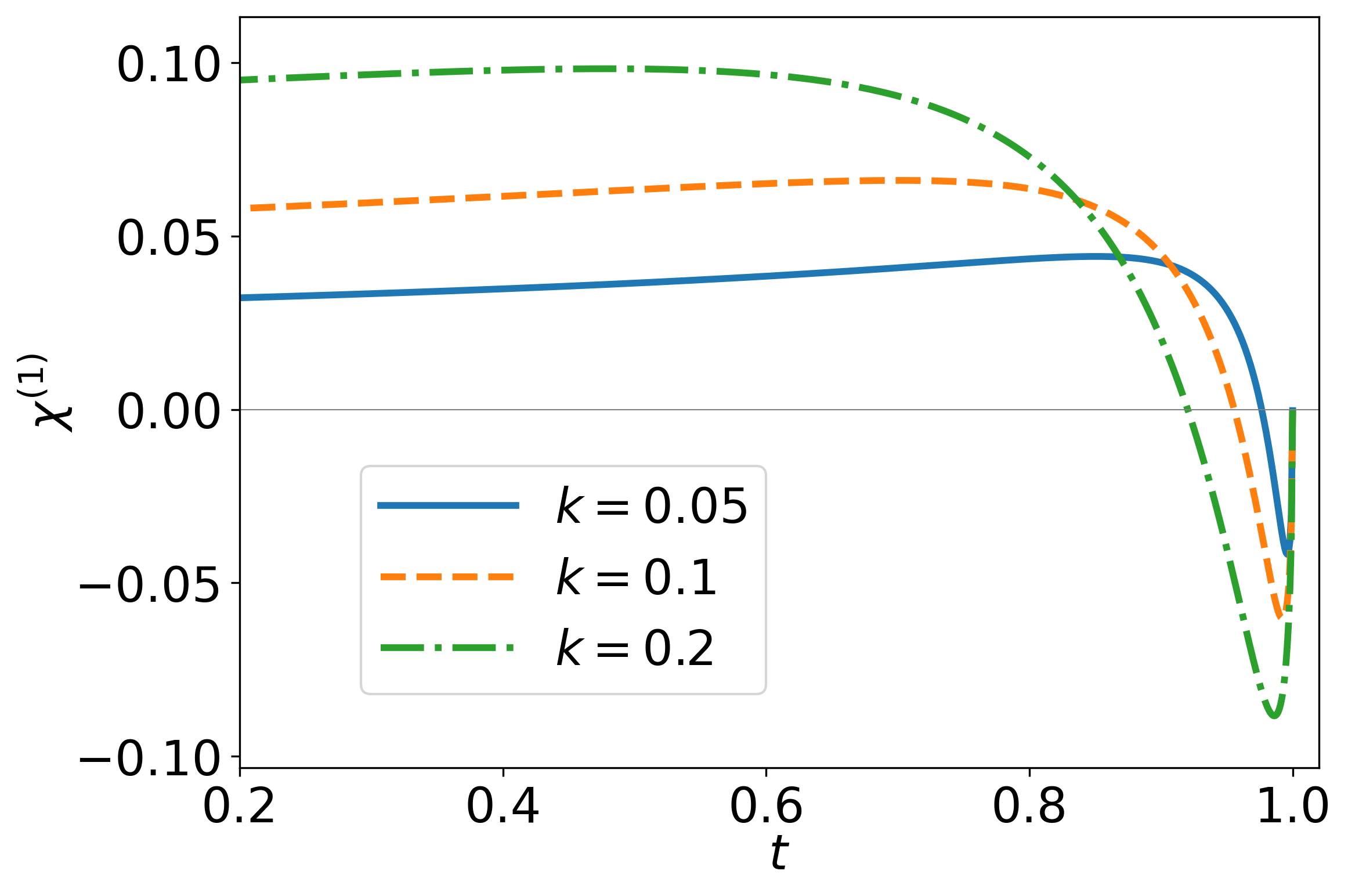}}\quad\quad
         \subfloat[]{\includegraphics[width=0.085\textwidth]{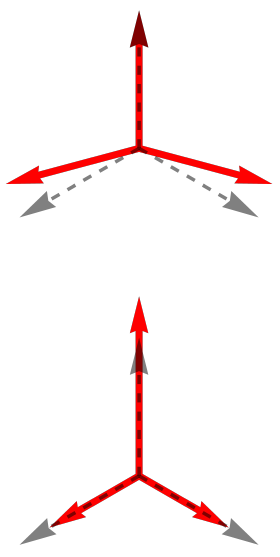}}
    \caption{(a) Temperature dependence of $\chi^{(1)}$ (in units of $M^3/(JN_c)^2$) for the toy model from scf calculations. $k\equiv K/(JN_c)$, $t \equiv T/T_c$. (b) Transverse (top) and longitudinal (bottom) local spin responses that lead to $\chi^{(1)}$. }
    \label{fig:mft}
\end{figure}

We next consider the $T\approx T_c$ case. After some algebra \cite{supp}, we obtain 
\begin{eqnarray}
    \chi^{(1)}\approx -\frac{4M^3}{5(JN_c)^2}\left(1+\frac{K}{JN_c}\right) n 
\end{eqnarray}
Surprisingly, $\chi^{(1)}$ is negative and approaches zero from below. Therefore it must have at least a sign change as well as two extrema between $T=0$ and $T_c$. This is verified by explicitly calculating $\chi^{(1)}$ from the self-consistently solved mean-field equations (Fig.~\ref{fig:mft}).

The sign change of $\chi^{(1)}$ versus temperature is a consequence of two competing mechanisms as illustrated in Fig.~\ref{fig:mft} (b) and detailed in \cite{supp}. At low temperatures, the ordered spins are saturated. The spins 2 and 3 can therefore only respond to the applied field by canting towards it. The $1/n$ dependence of $\chi^{(1)}$ at low temperatures in Eq.~\eqref{eq:chi0chi1lowT} is because the canting angle is determined by the ratio between the field-induced and the ordered spins. When the ordered spins shrink, the canting angle must increase, leading to the initial increase of $\chi^{(1)}$ with increasing $T$.

At higher temperatures, the ordered spins become much reduced from the saturated value, making them able to respond to the longitudinal field components, i.e., that along the local spin directions. For individual spins in a mean-field description, such a longitudinal response naturally has a nonzero and negative $\chi^{(1)}$, since increasing the spin is more difficult than suppressing it by the field, as reflected by the concave shape of the Langevin function. As $T$ approaches $T_c$ from below, the longitudinal mechanism becomes the dominant contribution to $\chi^{(1)}$, leading to its sign change. The above physical picture of two competing channels for $\chi^{(1)}$ is general and should hold beyond the mean-field treatment considered here. Though the sign change additionally requires the two contributions to have opposite signs which can depend on material-specific details. 

\textit{Discussion.---}Our work shows that certain AFM with strictly forbidden net magnetization can still be deterministically switched by uniform, static magnetic fields through symmetry-allowed nonlinear magnetic susceptibilities. Both collinear and noncollinear AFM can potentially be such nonlinear chiral AFM. However, the former may suffer from spin-flop transitions, unless $\chi^{(1)}$ has a large component perpendicular to the collinear axis. We also note that $\chi^{(1)}$ may be relevant in switching between near-degenerate states with the same net magnetization as in HoAgGe \cite{Zhao2024}.

Specifically for noncollinear AFM, our minimal model shows that large $\chi^{(1)}$ usually requires large anisotropy. HoAgGe and similar Ising-spin frustrated magnets are an extreme example, although in the Ising limit the minimal model does not directly apply and the discussion in \cite{Zhao2026} is more appropriate. Since easy-axis anisotropy scales quadratically with spin-orbit coupling \cite{Gay1986, Bruno1989, Chen2020}, compounds with heavier elements are more desirable. This is also consistent with the trend of Mn$_3X$N shown in Table~\ref{tab:Mn3XNchi01}: $X=$ Ag and Sn are in Row 5 of the periodic table and have much larger $\chi^{(1)}$ than $X=$ Ni, Ga, and Zn which are in Row 4.

In AFM with a symmetry-allowed net magnetization such as Mn$_3$Sn and the $\Gamma_{4g}$ phase of Mn$_3X$N, $\chi^{(1)}$ may still be relevant if the net magnetization $M$ is small. The critical field above which $\chi^{(1)}$ becomes more important than $M$ in separating the two TRS partners is $B_c = 6M/\chi^{(1)}$. For $M\sim 10^{-3}~\mu_B$ and $\chi^{(1)}\sim 10^{-4}~\mu_B \rm T^{-2}$ per unit cell, the latter is more important for deterministic switching if the coercive field is greater than 1 T. 

Together with \cite{Zhao2026}, the present work establishes a unified origin of nonlinear susceptibility: there is asymmetry between configurations reached by or under opposite magnetic fields. Such configurations can be thermally populated excited states as in HoAgGe, finite-field ground states such as for the 3-sublattice toy model at zero temperature, or scf corrections beyond linear spin-wave theories as indicated by the toy model. In this context, for large $\chi^{(1)}$, it is generally desirable if the system has nearby asymmetric instabilities induced by opposite magnetic fields. HoAgGe is again a good example, particularly since the important 1-spin-flip excitations contributing to $\chi^{(1)}$ of a given ground state are indeed the adjacent finite-field plateau states. Mn$_3$AgN and Mn$_3$SnN, which have larger $\chi^{(1)}$ than the other Mn$_3X$N, are also found in our DFT calculations to be more prone to have field-induced instabilities of flipping certain Mn spins. Our work shows that geometrically frustrated AFM with strong anisotropy are promising candidates for field-switchable chiral AFM through nonlinear susceptibilities.

\begin{acknowledgements} 

HC especially thanks Kan Zhao for introducing him to the Mn$_3$NiN and HoAgGe families, and for numerous illuminating discussions through the collaborations leading to Refs.~\cite{Zhao2019,Zhao2020,Zhao2024,Zhao2026}, which provided important inspiration for the present work. HC acknowledges support by NSF grant DMR-2531960. This work used Bridges-2 HPC at Pittsburgh Supercomputing Center through allocation PHY260075 from the Advanced Cyberinfrastructure Coordination Ecosystem: Services \& Support (ACCESS) program \cite{Boerner2023}, which is supported by U.S. National Science Foundation grants \#2138259, \#2138286, \#2138307, \#2137603, and \#2138296. PG acknowledges support by the German Research Foundation (DFG) through TRR360 (Project No. 492547816).

\end{acknowledgements} 
\appendix

\section{End Matter: Comparison with experimental $\chi^{(0)}$}
In this appendix we give a rough comparison between the DFT $\chi^{(0)}$ and experimental values reported in existing literature. $\chi^{(1)}$ is normally not measured experimentally and is not discussed here. The experimental numbers quoted below are mostly obtained by visually inspecting the relevant plots in the respective papers and should only be understood as order-of-magnitude estimates. 

\paragraph{Mn$_3$NiN ---}
Ref.~\cite{Zhao2019} reported (Fig.~2 (a) therein) the $\chi^{(0)}$ of stoichiometric Mn$_3$NiN at low temperatures near 0 K to be $\sim 0.02~\rm emu/mol/Oe$ in Gaussian units. To convert it to the proper units used in the main text, we use the following:
\begin{eqnarray}
    1~{\rm emu} &=&  \frac{1}{9.2740100783\times 10^{-21}}~\mu_B\\\nonumber
    1~{\rm mol} &=& 6.02214076\times 10^{23}~{\rm f.u.}\\\nonumber
    1~{\rm Oe} &=& \frac{10^{3}}{4\pi}~{\rm A/m} 
\end{eqnarray}
Assuming $\mu\approx \mu_0$, we have $1~{\rm Oe}\sim 10^{-4}~{\rm T}$. Therefore
\begin{eqnarray}
    1~\frac{\rm emu}{\rm mol\cdot Oe} \approx 1.79052972~\mu_B/{\rm T/f.u.}
\end{eqnarray}
The reported $\chi^{(0)}$ is therefore
\begin{eqnarray}
    \chi^{(0)}\approx 3.6\times 10^{-2}~\mu_B/{\rm T/f.u.}
\end{eqnarray}
Note that this value may be affected by the small net magnetization near zero field as shown in Fig.~S2 (a) in \cite{Zhao2019}.

Ref.~\cite{phdthesis} also contains useful magnetometry data of several Mn$_3X$N compounds. For Mn$_3$NiN at low temperatures $<100$ K, Fig.~104 for $M(H)$ at high fields $\sim 10$ T gives 
\begin{eqnarray}
    \chi^{(0)}\approx 8\times 10^{-3}~\mu_B/\rm T/f.u.
\end{eqnarray}

\paragraph{Mn$_3$AgN ---}
Ref.~\cite{Miao2024} gives the $M(H)$ hysteresis. For the $\Gamma_{5g}$ phase at 200 K, the magnetization is about $2.5~\rm emu/g$ at 40 kOe. The molar mass of Mn$_3$AgN is $286.689~{\rm g/mol}$. The $\chi^{(0)}$ is therefore
\begin{eqnarray}
    \chi^{(0)} &\approx & \frac{2.5}{40\times 10^3} \times {286.689} \frac{\rm emu}{\rm mol\cdot Oe} \approx 0.018 \frac{\rm emu}{\rm mol\cdot Oe} \\\nonumber
    &\approx & 3.2\times 10^{-2}~\mu_B/{\rm T/f.u.}
\end{eqnarray}

As another data point, Fig.~106 in \cite{phdthesis} gives $M(H)$ at 297 K, which at $10$ T is about 0.2 $\mu_B/\rm f.u.$ Therefore
\begin{eqnarray}
    \chi^{(0)}\approx 2\times 10^{-2}~\mu_B/{\rm T/f.u.}
\end{eqnarray}

\paragraph{Mn$_3$GaN ---}
Ref.~\cite{Sakakibara2015} gives the susceptibility of Mn$_3$GaN film at 4 K to be $6.7\times 10^{-4}~\rm \frac{emu}{cm^3\cdot Oe}$. They also give the volume of the unit cell for their film $V_{\rm uc} = 0.05836~\rm nm^3$. $\chi^{(0)}$ is therefore
\begin{eqnarray}
    \chi^{(0)}\approx 0.024 \frac{\rm emu}{\rm mol\cdot Oe} \approx 4.2\times 10^{-2}~\mu_B/{\rm T/f.u.}
\end{eqnarray}

Separately, Fig.~104 of \cite{phdthesis} gives $M(H)$ at 77 K. At $10$ T the magnetization is about $0.14~\mu_B/\rm f.u.$ Therefore
\begin{eqnarray}
    \chi^{(0)}\approx 1.4\times 10^{-2}~\mu_B/{\rm T/f.u.}
\end{eqnarray}

\paragraph{Mn$_3$ZnN ---}
Ref.~\cite{Sun2012} reported the $1/\chi^{(0)}$ near 0 K to be $\sim 50~\rm Oe \cdot mol/emu$. We therefore get
\begin{eqnarray}
    \chi^{(0)}\approx 0.02 \frac{\rm emu}{\rm mol\cdot Oe} 
    \approx 3.6\times 10^{-2}~\mu_B/{\rm T/f.u.}
\end{eqnarray}

\paragraph{Mn$_3$SnN ---}
Fig.~105 of \cite{phdthesis} gives $M(H)$ at 293 K when Mn$_3$SnN should be in the $\Gamma_{5g}$ phase. At 10 T the magnetization is about $0.09~\mu_B/\rm f.u.$ Consequently
\begin{eqnarray}
    \chi^{(0)}\approx 9\times 10^{-3} ~\mu_B/{\rm T/f.u.}
\end{eqnarray}

Compared to Table~I in the main text, the above experimental $\chi^{(0)}$ are roughly one order of magnitude larger, though the differences for some materials are smaller than the others. Such a discrepancy can be due to any of the following reasons, which is not an exclusive list. (1) The DFT calculations only considered $\chi^{(0)}$ for the stoichiometric, unstrained Mn$_3X$N, in pure $\Gamma_{5g}$ phase, at zero temperature, for fields along $[1\bar{1}0]$, which can be different from the conditions in the above experiments. (2) The DFT calculations do not consider orbital contributions to $\chi^{(0)}$, including the orbital-orbital and spin-orbital suscetpibilities. However, for noncollinear AFM such orbital contributions to magnetism can be significant \cite{Chen2020}. (3) The calculated $\chi^{(0)}$ may be sensitive to other DFT-specific issues such as correlation effects, pseudopotentials, etc., that are not explored here.

\bibliography{nlcafm}

@Article{Giordano1980,
  author           = {Giordano, N. and Wolf, W. P.},
  journal          = {Physical Review B},
  title            = {{Induced staggered magnetic fields in antiferromagnets: Microscopic mechanisms}},
  year             = {1980},
  issn             = {0163-1829},
  month            = mar,
  number           = {5},
  pages            = {2008--2026},
  volume           = {21},
  creationdate     = {2025-01-29T10:58:48},
  doi              = {10.1103/physrevb.21.2008},
  modificationdate = {2025-02-28T10:24:36},
  publisher        = {American Physical Society (APS)},
}

@Article{Gorodetsky1967,
  author           = {Gorodetsky, G. and Sharon, B. and Shtrikman, S.},
  journal          = {Solid State Communications},
  title            = {{Linear effect of the magnetic field on the magnetic susceptibility in antiferromagnetic DyFeO3}},
  year             = {1967},
  issn             = {0038-1098},
  month            = sep,
  number           = {9},
  pages            = {739--741},
  volume           = {5},
  creationdate     = {2025-02-04T15:06:14},
  doi              = {10.1016/0038-1098(67)90362-6},
  modificationdate = {2025-02-28T10:24:36},
  publisher        = {Elsevier BV},
}

@Article{Kharchenko1995,
  author           = {Kharchenko, N.F. and Szymczak, R. and Baran, M.},
  journal          = {Journal of Magnetism and Magnetic Materials},
  title            = {{Quadratic in field contribution to the magnetization of antiferromagnetic CoF2}},
  year             = {1995},
  issn             = {0304-8853},
  month            = feb,
  pages            = {161--162},
  volume           = {140–144},
  creationdate     = {2025-02-04T15:07:03},
  doi              = {10.1016/0304-8853(94)01006-4},
  modificationdate = {2025-02-28T10:24:36},
  publisher        = {Elsevier BV},
}

@Article{Gregg1990,
  author           = {Gregg, J. F. and Kob, W. and Lord, J. S. and Morris, I. D. and Pfeffer, J. Z. and Schilling, R. and Wells, M. R. and Wolf, W. P.},
  journal          = {Journal of Applied Physics},
  title            = {{Relaxation dynamics of metastable antiferromagnetic states}},
  year             = {1990},
  issn             = {1089-7550},
  month            = may,
  number           = {9},
  pages            = {5430--5432},
  volume           = {67},
  creationdate     = {2025-02-04T15:08:31},
  doi              = {10.1063/1.344579},
  modificationdate = {2025-02-28T10:24:36},
  publisher        = {AIP Publishing},
}

@Article{Wolf1990,
  author           = {Wolf, W.P.},
  journal          = {Journal of Magnetism and Magnetic Materials},
  title            = {{Induced staggered field effects in antiferromagnets}},
  year             = {1990},
  issn             = {0304-8853},
  month            = dec,
  pages            = {197--198},
  volume           = {90–91},
  creationdate     = {2025-02-04T15:09:04},
  doi              = {10.1016/s0304-8853(10)80069-1},
  modificationdate = {2025-02-28T10:24:36},
  publisher        = {Elsevier BV},
}

@Article{Alben1975,
  author           = {Alben, R. and Blume, M. and Corliss, L. M. and Hastings, J. M.},
  journal          = {Physical Review B},
  title            = {{Induced staggered magnetic fields in antiferromagnets}},
  year             = {1975},
  issn             = {0556-2805},
  month            = jan,
  number           = {1},
  pages            = {295--299},
  volume           = {11},
  creationdate     = {2025-02-04T15:15:22},
  doi              = {10.1103/physrevb.11.295},
  modificationdate = {2025-02-28T10:24:36},
  publisher        = {American Physical Society (APS)},
}

@Article{Foglio1977,
  author           = {Foglio, M. E. and Blume, M.},
  journal          = {Physical Review B},
  title            = {{Induced staggered magnetic fields in antiferromagnets: Theg-factor mechanism}},
  year             = {1977},
  issn             = {0556-2805},
  month            = apr,
  number           = {7},
  pages            = {3465--3469},
  volume           = {15},
  creationdate     = {2025-02-04T15:15:55},
  doi              = {10.1103/physrevb.15.3465},
  modificationdate = {2025-02-28T10:24:36},
  publisher        = {American Physical Society (APS)},
}

@Article{Mukamel1977,
  author           = {Mukamel, D. and Blume, M.},
  journal          = {Physical Review B},
  title            = {{Induced staggered magnetic fields in dysprosium aluminum garnet}},
  year             = {1977},
  issn             = {0556-2805},
  month            = may,
  number           = {9},
  pages            = {4516--4523},
  volume           = {15},
  creationdate     = {2025-02-04T15:16:20},
  doi              = {10.1103/physrevb.15.4516},
  modificationdate = {2025-02-28T10:24:36},
  publisher        = {American Physical Society (APS)},
}

@Article{Blume1974,
  author           = {Blume, M. and Corliss, L. M. and Hastings, J. M. and Schiller, E.},
  journal          = {Physical Review Letters},
  title            = {{Observation of an Antiferromagnet in an Induced Staggered Magnetic Field: Dysprosium Aluminum Garnet near the Tricritical Point}},
  year             = {1974},
  issn             = {0031-9007},
  month            = mar,
  number           = {10},
  pages            = {544--547},
  volume           = {32},
  creationdate     = {2025-02-04T15:17:00},
  doi              = {10.1103/physrevlett.32.544},
  modificationdate = {2025-02-28T10:24:36},
  publisher        = {American Physical Society (APS)},
}

@Article{Dillon1974,
  author           = {Dillon, J. F. and Chen, E. Yi and Giordano, N. and Wolf, W. P.},
  journal          = {Physical Review Letters},
  title            = {{Time-Reversed Antiferromagnetic States in Dysprosium Aluminum Garnet}},
  year             = {1974},
  issn             = {0031-9007},
  month            = jul,
  number           = {2},
  pages            = {98--101},
  volume           = {33},
  creationdate     = {2025-02-04T15:17:15},
  doi              = {10.1103/physrevlett.33.98},
  modificationdate = {2025-02-28T10:24:36},
  publisher        = {American Physical Society (APS)},
}

@Article{Fujita2015,
  author           = {Fujita, T. C. and Kozuka, Y. and Uchida, M. and Tsukazaki, A. and Arima, T. and Kawasaki, M.},
  journal          = {Scientific Reports},
  title            = {{Odd-parity magnetoresistance in pyrochlore iridate thin films with broken time-reversal symmetry}},
  year             = {2015},
  issn             = {2045-2322},
  month            = may,
  number           = {1},
  volume           = {5},
  creationdate     = {2025-02-04T15:18:38},
  doi              = {10.1038/srep09711},
  modificationdate = {2025-02-28T10:24:36},
  publisher        = {Springer Science and Business Media LLC},
}

@Article{Aroyo2006,
  author           = {Aroyo, Mois Ilia and Perez-Mato, Juan Manuel and Capillas, Cesar and Kroumova, Eli and Ivantchev, Svetoslav and Madariaga, Gotzon and Kirov, Asen and Wondratschek, Hans},
  journal          = {Zeitschrift für Kristallographie - Crystalline Materials},
  title            = {{Bilbao Crystallographic Server: I. Databases and crystallographic computing programs}},
  year             = {2006},
  issn             = {2194-4946},
  month            = jan,
  number           = {1},
  pages            = {15--27},
  volume           = {221},
  creationdate     = {2025-02-28T10:25:02},
  doi              = {10.1524/zkri.2006.221.1.15},
  modificationdate = {2025-02-28T10:25:02},
  publisher        = {Walter de Gruyter GmbH},
}

@Article{Aroyo2006a,
  author           = {Aroyo, Mois I. and Kirov, Asen and Capillas, Cesar and Perez-Mato, J. M. and Wondratschek, Hans},
  journal          = {Acta Crystallographica Section A Foundations of Crystallography},
  title            = {{Bilbao Crystallographic Server. II. Representations of crystallographic point groups and space groups}},
  year             = {2006},
  issn             = {0108-7673},
  month            = mar,
  number           = {2},
  pages            = {115--128},
  volume           = {62},
  creationdate     = {2025-02-28T10:25:37},
  doi              = {10.1107/s0108767305040286},
  modificationdate = {2025-02-28T10:25:37},
  publisher        = {International Union of Crystallography (IUCr)},
}

@Article{Gallego2019,
  author           = {Gallego, Samuel V. and Etxebarria, Jesus and Elcoro, Luis and Tasci, Emre S. and Perez-Mato, J. Manuel},
  journal          = {Acta Crystallographica Section A Foundations and Advances},
  title            = {{Automatic calculation of symmetry-adapted tensors in magnetic and non-magnetic materials: a new tool of the Bilbao Crystallographic Server}},
  year             = {2019},
  issn             = {2053-2733},
  month            = apr,
  number           = {3},
  pages            = {438--447},
  volume           = {75},
  creationdate     = {2025-02-28T10:26:28},
  doi              = {10.1107/s2053273319001748},
  modificationdate = {2025-02-28T10:26:28},
  publisher        = {International Union of Crystallography (IUCr)},
}

@Article{Zhao2020,
  author           = {Zhao, Kan and Deng, Hao and Chen, Hua and Ross, Kate A. and Petříček, Vaclav and Günther, Gerrit and Russina, Margarita and Hutanu, Vladimir and Gegenwart, Philipp},
  journal          = {Science},
  title            = {{Realization of the kagome spin ice state in a frustrated intermetallic compound}},
  year             = {2020},
  issn             = {1095-9203},
  month            = mar,
  number           = {6483},
  pages            = {1218--1223},
  volume           = {367},
  creationdate     = {2025-02-28T10:28:33},
  doi              = {10.1126/science.aaw1666},
  modificationdate = {2025-02-28T10:28:33},
  publisher        = {American Association for the Advancement of Science (AAAS)},
}

@Article{Zhao2024,
  author           = {Zhao, K. and Tokiwa, Y. and Chen, H. and Gegenwart, P.},
  journal          = {Nature Physics},
  title            = {{Discrete degeneracies distinguished by the anomalous Hall effect in a metallic kagome ice compound}},
  year             = {2024},
  issn             = {1745-2481},
  month            = jan,
  number           = {3},
  pages            = {442--449},
  volume           = {20},
  creationdate     = {2025-02-28T10:29:01},
  doi              = {10.1038/s41567-023-02307-w},
  modificationdate = {2025-02-28T10:29:01},
  publisher        = {Springer Science and Business Media LLC},
}

@Article{Chen2020,
  author           = {Chen, Hua and Wang, Tzu-Cheng and Xiao, Di and Guo, Guang-Yu and Niu, Qian and MacDonald, Allan H.},
  journal          = {Physical Review B},
  title            = {{Manipulating anomalous Hall antiferromagnets with magnetic fields}},
  year             = {2020},
  issn             = {2469-9969},
  month            = mar,
  number           = {10},
  pages            = {104418},
  volume           = {101},
  creationdate     = {2025-10-06T09:58:47},
  doi              = {10.1103/physrevb.101.104418},
  modificationdate = {2025-10-06T09:58:47},
  publisher        = {American Physical Society (APS)},
}

@Book{Altland2023,
  author           = {Altland, Alexander and Simons, Ben},
  publisher        = {Cambridge University Press},
  title            = {{Condensed Matter Field Theory}},
  year             = {2023},
  isbn             = {9781108494601},
  month            = aug,
  creationdate     = {2025-10-06T09:59:12},
  doi              = {10.1017/9781108781244},
  modificationdate = {2025-10-06T09:59:12},
}

@Article{Kondou2021,
  author           = {Kondou, Kouta and Chen, Hua and Tomita, Takahiro and Ikhlas, Muhammad and Higo, Tomoya and MacDonald, Allan H. and Nakatsuji, Satoru and Otani, YoshiChika},
  journal          = {Nature Communications},
  title            = {{Giant field-like torque by the out-of-plane magnetic spin Hall effect in a topological antiferromagnet}},
  year             = {2021},
  issn             = {2041-1723},
  month            = nov,
  number           = {1},
  volume           = {12},
  creationdate     = {2025-10-06T10:00:18},
  doi              = {10.1038/s41467-021-26453-y},
  modificationdate = {2026-08-07T00:38:57},
  publisher        = {Springer Science and Business Media LLC},
}

@Article{Chen2014,
  author           = {Chen, Hua and Niu, Qian and MacDonald, A. H.},
  journal          = {Phys. Rev. Lett.},
  title            = {{Anomalous Hall Effect Arising from Noncollinear Antiferromagnetism}},
  year             = {2014},
  month            = {Jan},
  pages            = {017205},
  volume           = {112},
  creationdate     = {2026-08-06T18:37:29},
  doi              = {10.1103/PhysRevLett.112.017205},
  issue            = {1},
  modificationdate = {2026-08-06T18:37:29},
  numpages         = {5},
  publisher        = {American Physical Society},
  url              = {https://link.aps.org/doi/10.1103/PhysRevLett.112.017205},
}

@Article{Nakatsuji2015,
  author           = {Satoru Nakatsuji and Naoki Kiyohara and Tomoya Higo},
  journal          = {Nature},
  title            = {{Large anomalous Hall effect in a non-collinear antiferromagnet at room temperature}},
  year             = {2015},
  pages            = {212-215},
  volume           = {527},
  creationdate     = {2026-08-06T18:37:29},
  modificationdate = {2026-08-06T18:37:29},
}

@Article{Kuebler2014,
  author           = {J. K\"{u}bler and C. Felser},
  journal          = {{EPL} (Europhysics Letters)},
  title            = {{Non-collinear antiferromagnets and the anomalous Hall effect}},
  year             = {2014},
  month            = {dec},
  number           = {6},
  pages            = {67001},
  volume           = {108},
  creationdate     = {2026-08-06T18:37:29},
  doi              = {10.1209/0295-5075/108/67001},
  modificationdate = {2026-08-06T18:37:29},
  publisher        = {{IOP} Publishing},
  url              = {https://doi.org/10.1209%2F0295-5075%2F108%2F67001},
}

@Article{Kimata2019,
  author           = {Kimata, Motoi and Chen, Hua and Kondou, Kouta and Sugimoto, Satoshi and Muduli, Prasanta K. and Ikhlas, Muhammad and Omori, Yasutomo and Tomita, Takahiro and MacDonald, Allan H. and Nakatsuji, Satoru and Otani, Yoshichika},
  journal          = {Nature},
  title            = {{Magnetic and magnetic-inverse spin Hall effects in a non-collinear antiferromagnet}},
  year             = {2019},
  issn             = {1476-4687},
  number           = {7741},
  pages            = {627-630},
  volume           = {565},
  creationdate     = {2026-08-06T18:37:29},
  doi              = {10.1038/s41586-018-0853-0},
  modificationdate = {2026-08-07T00:38:57},
  url              = {https://doi.org/10.1038/s41586-018-0853-0},
}

@Article{Tomizawa2009,
  author           = {Tomizawa, Takeshi and Kontani, Hiroshi},
  journal          = {Phys. Rev. B},
  title            = {{Anomalous Hall effect in the ${t}_{2g}$ orbital kagome lattice due to noncollinearity: Significance of the orbital Aharonov-Bohm effect}},
  year             = {2009},
  month            = {Sep},
  pages            = {100401},
  volume           = {80},
  creationdate     = {2026-08-06T18:37:29},
  doi              = {10.1103/PhysRevB.80.100401},
  issue            = {10},
  modificationdate = {2026-08-06T18:37:29},
  numpages         = {4},
  publisher        = {American Physical Society},
  url              = {https://link.aps.org/doi/10.1103/PhysRevB.80.100401},
}

@Article{Ohgushi2000,
  author           = {Ohgushi, Kenya and Murakami, Shuichi and Nagaosa, Naoto},
  journal          = {Phys. Rev. B},
  title            = {{Spin anisotropy and quantum Hall effect in the kagom\'e lattice: Chiral spin state based on a ferromagnet}},
  year             = {2000},
  month            = {Sep},
  pages            = {R6065--R6068},
  volume           = {62},
  creationdate     = {2026-08-06T18:37:29},
  doi              = {10.1103/PhysRevB.62.R6065},
  issue            = {10},
  modificationdate = {2026-08-06T18:37:29},
  numpages         = {0},
  publisher        = {American Physical Society},
  url              = {https://link.aps.org/doi/10.1103/PhysRevB.62.R6065},
}

@Article{Shindou2001,
  author           = {Shindou, Ryuichi and Nagaosa, Naoto},
  journal          = {Phys. Rev. Lett.},
  title            = {{Orbital Ferromagnetism and Anomalous Hall Effect in Antiferromagnets on the Distorted fcc Lattice}},
  year             = {2001},
  month            = {Aug},
  pages            = {116801},
  volume           = {87},
  creationdate     = {2026-08-06T18:37:30},
  doi              = {10.1103/PhysRevLett.87.116801},
  issue            = {11},
  modificationdate = {2026-08-06T18:37:30},
  numpages         = {4},
  publisher        = {American Physical Society},
  url              = {https://link.aps.org/doi/10.1103/PhysRevLett.87.116801},
}

@Article{Smejkal2020,
  author           = {{\v S}mejkal, Libor and Gonz{\'a}lez-Hern{\'a}ndez, Rafael and Jungwirth, T. and Sinova, J.},
  journal          = {Science Advances},
  title            = {{Crystal time-reversal symmetry breaking and spontaneous Hall effect in collinear antiferromagnets}},
  year             = {2020},
  number           = {23},
  pages            = {eaaz8809},
  volume           = {6},
  creationdate     = {2026-08-06T18:37:30},
  doi              = {10.1126/sciadv.aaz8809},
  modificationdate = {2026-08-06T18:37:30},
  publisher        = {American Association for the Advancement of Science},
  url              = {https://advances.sciencemag.org/content/6/23/eaaz8809},
}

@Article{Solovyev1997,
  author           = {Solovyev, I. V.},
  journal          = {Phys. Rev. B},
  title            = {{Magneto-optical effect in the weak ferromagnets ${\mathrm{LaMO}}_{3}$ (M= Cr, Mn, and Fe)}},
  year             = {1997},
  month            = {Apr},
  pages            = {8060--8063},
  volume           = {55},
  creationdate     = {2026-08-06T18:37:30},
  doi              = {10.1103/PhysRevB.55.8060},
  issue            = {13},
  modificationdate = {2026-08-06T18:37:30},
  numpages         = {0},
  publisher        = {American Physical Society},
  url              = {https://link.aps.org/doi/10.1103/PhysRevB.55.8060},
}

@Article{Nayak2016,
  author           = {Nayak, Ajaya K. and Fischer, Julia Erika and Sun, Yan and Yan, Binghai and Karel, Julie and Komarek, Alexander C. and Shekhar, Chandra and Kumar, Nitesh and Schnelle, Walter and K{\"u}bler, J{\"u}rgen and Felser, Claudia and Parkin, Stuart S. P.},
  journal          = {Science Advances},
  title            = {{Large anomalous Hall effect driven by a nonvanishing Berry curvature in the noncolinear antiferromagnet Mn$_3$Ge}},
  year             = {2016},
  number           = {4},
  pages            = {e1501870},
  volume           = {2},
  creationdate     = {2026-08-06T18:37:30},
  doi              = {10.1126/sciadv.1501870},
  modificationdate = {2026-08-06T18:37:30},
  publisher        = {American Association for the Advancement of Science},
  url              = {https://advances.sciencemag.org/content/2/4/e1501870},
}

@Article{Zhou2019,
  author           = {Zhou, Xiaodong and Hanke, Jan-Philipp and Feng, Wanxiang and Li, Fei and Guo, Guang-Yu and Yao, Yugui and Bl\"ugel, Stefan and Mokrousov, Yuriy},
  journal          = {Phys. Rev. B},
  title            = {{Spin-order dependent anomalous Hall effect and magneto-optical effect in the noncollinear antiferromagnets ${\mathrm{Mn}}_{3}X\mathrm{N}$ with $X=\mathrm{Ga}$, Zn, Ag, or Ni}},
  year             = {2019},
  month            = {Mar},
  pages            = {104428},
  volume           = {99},
  creationdate     = {2026-08-06T18:37:30},
  doi              = {10.1103/PhysRevB.99.104428},
  issue            = {10},
  modificationdate = {2026-08-06T19:46:33},
  numpages         = {13},
  publisher        = {American Physical Society},
  url              = {https://link.aps.org/doi/10.1103/PhysRevB.99.104428},
}

@Article{Gurung2019,
  author           = {Gurung, Gautam and Shao, Ding-Fu and Paudel, Tula R. and Tsymbal, Evgeny Y.},
  journal          = {Phys. Rev. Materials},
  title            = {{Anomalous Hall conductivity of noncollinear magnetic antiperovskites}},
  year             = {2019},
  month            = {Apr},
  pages            = {044409},
  volume           = {3},
  creationdate     = {2026-08-06T18:37:30},
  doi              = {10.1103/PhysRevMaterials.3.044409},
  issue            = {4},
  modificationdate = {2026-08-06T19:46:44},
  numpages         = {8},
  publisher        = {American Physical Society},
  url              = {https://link.aps.org/doi/10.1103/PhysRevMaterials.3.044409},
}

@Article{Boldrin2019,
  author           = {Boldrin, David and Samathrakis, Ilias and Zemen, Jan and Mihai, Andrei and Zou, Bin and Johnson, Freya and Esser, Bryan D. and McComb, David W. and Petrov, Peter K. and Zhang, Hongbin and Cohen, Lesley F.},
  journal          = {Phys. Rev. Materials},
  title            = {{Anomalous Hall effect in noncollinear antiferromagnetic ${\mathrm{Mn}}_{3}\mathrm{NiN}$ thin films}},
  year             = {2019},
  month            = {Sep},
  pages            = {094409},
  volume           = {3},
  creationdate     = {2026-08-06T18:37:30},
  doi              = {10.1103/PhysRevMaterials.3.094409},
  issue            = {9},
  modificationdate = {2026-08-06T19:47:06},
  numpages         = {6},
  publisher        = {American Physical Society},
  url              = {https://link.aps.org/doi/10.1103/PhysRevMaterials.3.094409},
}

@Article{Zhao2019,
  author           = {Zhao, K. and Hajiri, T. and Chen, H. and Miki, R. and Asano, H. and Gegenwart, P.},
  journal          = {Phys. Rev. B},
  title            = {{Anomalous Hall effect in the noncollinear antiferromagnetic antiperovskite ${\mathrm{Mn}}_{3}{\mathrm{Ni}}_{1\ensuremath{-}x}{\mathrm{Cu}}_{x}\mathrm{N}$}},
  year             = {2019},
  month            = {Jul},
  pages            = {045109},
  volume           = {100},
  creationdate     = {2026-08-06T18:37:30},
  doi              = {10.1103/PhysRevB.100.045109},
  issue            = {4},
  modificationdate = {2026-08-06T19:42:15},
  numpages         = {6},
  publisher        = {American Physical Society},
  url              = {https://link.aps.org/doi/10.1103/PhysRevB.100.045109},
}

@Article{Liu2018,
  author           = {Liu, Z. Q. and Chen, H. and Wang, J. M. and Liu, J. H. and Wang, K. and Feng, Z. X. and Yan, H. and Wang, X. R. and Jiang, C. B. and Coey, J. M. D. and MacDonald, A. H.},
  journal          = {Nature Electronics},
  title            = {{Electrical switching of the topological anomalous Hall effect in a non-collinear antiferromagnet above room temperature}},
  year             = {2018},
  issn             = {2520-1131},
  month            = {Mar},
  number           = {3},
  pages            = {172-177},
  volume           = {1},
  creationdate     = {2026-08-06T18:37:30},
  day              = {01},
  doi              = {10.1038/s41928-018-0040-1},
  modificationdate = {2026-08-06T18:37:30},
  url              = {https://doi.org/10.1038/s41928-018-0040-1},
}

@Article{Chen2022,
  author           = {Chen, Hua},
  journal          = {Phys. Rev. B},
  title            = {{Electronic chiralization as an indicator of the anomalous Hall effect in unconventional magnetic systems}},
  year             = {2022},
  month            = {Jul},
  pages            = {024421},
  volume           = {106},
  creationdate     = {2026-08-06T18:37:30},
  doi              = {10.1103/PhysRevB.106.024421},
  issue            = {2},
  modificationdate = {2026-08-06T18:37:30},
  numpages         = {13},
  publisher        = {American Physical Society},
  url              = {https://link.aps.org/doi/10.1103/PhysRevB.106.024421},
}

@Misc{supp,
  title            = {{Supplemental Material}},
  creationdate     = {2026-08-06T18:37:31},
  modificationdate = {2026-08-07T01:39:11},
}

@Article{Liu2026,
  author           = {Liu, Zheng and Gao, Yang and Niu, Qian},
  journal          = {Physical Review Letters},
  title            = {{Rigid-Body Anisotropy in Noncollinear Antiferromagnets}},
  year             = {2026},
  issn             = {1079-7114},
  month            = jan,
  number           = {2},
  volume           = {136},
  creationdate     = {2026-08-06T18:37:32},
  doi              = {10.1103/64wt-51gd},
  modificationdate = {2026-08-06T18:37:32},
  publisher        = {American Physical Society (APS)},
}

@Article{Liu2025,
  author           = {Liu, Zheng and Wei, Mengjie and Peng, Wenzhi and Hou, Dazhi and Gao, Yang and Niu, Qian},
  journal          = {Physical Review X},
  title            = {{Multipolar Anisotropy in Anomalous Hall Effect from Spin-Group Symmetry Breaking}},
  year             = {2025},
  issn             = {2160-3308},
  month            = jul,
  number           = {3},
  pages            = {031006},
  volume           = {15},
  creationdate     = {2026-08-06T18:37:32},
  doi              = {10.1103/physrevx.15.031006},
  modificationdate = {2026-08-06T18:37:32},
  publisher        = {American Physical Society (APS)},
}

@Article{Jungwirth2026,
  author           = {Jungwirth, Tomas and Sinova, Jairo and Fernandes, Rafael M. and Liu, Qihang and Watanabe, Hikaru and Murakami, Shuichi and Nakatsuji, Satoru and Šmejkal, Libor},
  journal          = {Nature},
  title            = {{Symmetry, microscopy and spectroscopy signatures of altermagnetism}},
  year             = {2026},
  issn             = {1476-4687},
  month            = jan,
  number           = {8098},
  pages            = {837--847},
  volume           = {649},
  creationdate     = {2026-08-06T18:37:32},
  doi              = {10.1038/s41586-025-09883-2},
  modificationdate = {2026-08-06T18:37:32},
  publisher        = {Springer Science and Business Media LLC},
}

@Article{Song2025,
  author           = {Song, Cheng and Bai, Hua and Zhou, Zhiyuan and Han, Lei and Reichlova, Helena and Dil, J. Hugo and Liu, Junwei and Chen, Xianzhe and Pan, Feng},
  journal          = {Nature Reviews Materials},
  title            = {{Altermagnets as a new class of functional materials}},
  year             = {2025},
  issn             = {2058-8437},
  month            = feb,
  number           = {6},
  pages            = {473--485},
  volume           = {10},
  creationdate     = {2026-08-06T18:37:32},
  doi              = {10.1038/s41578-025-00779-1},
  modificationdate = {2026-08-06T18:37:32},
  publisher        = {Springer Science and Business Media LLC},
}

@Article{Tamang2025,
  author           = {Tamang, Rupam and Gurung, Shivraj and Rai, Dibya Prakash and Brahimi, Samy and Lounis, Samir},
  journal          = {Magnetism},
  title            = {{Altermagnetism and Altermagnets: A Brief Review}},
  year             = {2025},
  issn             = {2673-8724},
  month            = jul,
  number           = {3},
  pages            = {17},
  volume           = {5},
  creationdate     = {2026-08-06T18:37:32},
  doi              = {10.3390/magnetism5030017},
  modificationdate = {2026-08-06T18:37:32},
  publisher        = {MDPI AG},
}

@Article{Shim2025,
  author           = {Shim, Soho and Mehraeen, M. and Sklenar, Joseph and Zhang, Steven S.-L. and Hoffmann, Axel and Mason, Nadya},
  journal          = {Annual Review of Condensed Matter Physics},
  title            = {{Spin-Polarized Antiferromagnetic Metals}},
  year             = {2025},
  issn             = {1947-5462},
  month            = mar,
  number           = {1},
  pages            = {103--120},
  volume           = {16},
  creationdate     = {2026-08-06T18:37:32},
  doi              = {10.1146/annurev-conmatphys-042924-123620},
  modificationdate = {2026-08-06T18:37:32},
  publisher        = {Annual Reviews},
}

@Article{Guo2025,
  author           = {Guo, Zhenzhou and Wang, Xiaotian and Wang, Wenhong and Zhang, Gang and Zhou, Xiaodong and Cheng, Zhenxiang},
  journal          = {Advanced Materials},
  title            = {{Spin‐Polarized Antiferromagnets for Spintronics}},
  year             = {2025},
  issn             = {1521-4095},
  month            = jun,
  number           = {36},
  volume           = {37},
  creationdate     = {2026-08-06T18:37:32},
  doi              = {10.1002/adma.202505779},
  modificationdate = {2026-08-06T18:37:32},
  publisher        = {Wiley},
}

@Article{Smejkal2022,
  author           = {Šmejkal, Libor and Sinova, Jairo and Jungwirth, Tomas},
  journal          = {Physical Review X},
  title            = {{Beyond Conventional Ferromagnetism and Antiferromagnetism: A Phase with Nonrelativistic Spin and Crystal Rotation Symmetry}},
  year             = {2022},
  issn             = {2160-3308},
  month            = sep,
  number           = {3},
  pages            = {031042},
  volume           = {12},
  creationdate     = {2026-08-06T18:37:32},
  doi              = {10.1103/physrevx.12.031042},
  modificationdate = {2026-08-06T18:37:32},
  publisher        = {American Physical Society (APS)},
}

@Article{Smejkal2022a,
  author           = {Šmejkal, Libor and Sinova, Jairo and Jungwirth, Tomas},
  journal          = {Physical Review X},
  title            = {{Emerging Research Landscape of Altermagnetism}},
  year             = {2022},
  issn             = {2160-3308},
  month            = dec,
  number           = {4},
  pages            = {040501},
  volume           = {12},
  creationdate     = {2026-08-06T18:37:32},
  doi              = {10.1103/physrevx.12.040501},
  modificationdate = {2026-08-06T18:37:32},
  publisher        = {American Physical Society (APS)},
}

@Article{Liu2022,
  author           = {Liu, Pengfei and Li, Jiayu and Han, Jingzhi and Wan, Xiangang and Liu, Qihang},
  journal          = {Physical Review X},
  title            = {{Spin-Group Symmetry in Magnetic Materials with Negligible Spin-Orbit Coupling}},
  year             = {2022},
  issn             = {2160-3308},
  month            = apr,
  number           = {2},
  pages            = {021016},
  volume           = {12},
  creationdate     = {2026-08-06T18:37:32},
  doi              = {10.1103/physrevx.12.021016},
  modificationdate = {2026-08-06T18:37:32},
  publisher        = {American Physical Society (APS)},
}

@Article{Brinkman1966,
  author           = {W. F. Brinkman and R. J. Elliott},
  journal          = {Proceedings of the Royal Society of London. Series A, Mathematical and Physical Sciences},
  title            = {{Theory of Spin-Space Groups}},
  year             = {1966},
  issn             = {00804630},
  number           = {1438},
  pages            = {343--358},
  volume           = {294},
  creationdate     = {2026-08-06T18:37:32},
  modificationdate = {2026-08-06T18:37:32},
  publisher        = {The Royal Society},
  url              = {http://www.jstor.org/stable/2415409},
  urldate          = {2026-03-09},
}

@Article{Litvin1974,
  author           = {Litvin, D.B. and Opechowski, W.},
  journal          = {Physica},
  title            = {{Spin groups}},
  year             = {1974},
  issn             = {0031-8914},
  month            = sep,
  number           = {3},
  pages            = {538--554},
  volume           = {76},
  creationdate     = {2026-08-06T18:37:32},
  doi              = {10.1016/0031-8914(74)90157-8},
  modificationdate = {2026-08-06T18:37:32},
  publisher        = {Elsevier BV},
}

@Article{Litvin1977,
  author           = {Litvin, D. B.},
  journal          = {Acta Crystallographica Section A},
  title            = {{Spin point groups}},
  year             = {1977},
  issn             = {0567-7394},
  month            = mar,
  number           = {2},
  pages            = {279--287},
  volume           = {33},
  creationdate     = {2026-08-06T18:37:32},
  doi              = {10.1107/s0567739477000709},
  modificationdate = {2026-08-06T18:37:32},
  publisher        = {International Union of Crystallography (IUCr)},
}

@Article{Chen2024,
  author           = {Chen, Xiaobing and Ren, Jun and Zhu, Yanzhou and Yu, Yutong and Zhang, Ao and Liu, Pengfei and Li, Jiayu and Liu, Yuntian and Li, Caiheng and Liu, Qihang},
  journal          = {Physical Review X},
  title            = {{Enumeration and Representation Theory of Spin Space Groups}},
  year             = {2024},
  issn             = {2160-3308},
  month            = aug,
  number           = {3},
  pages            = {031038},
  volume           = {14},
  creationdate     = {2026-08-06T18:37:32},
  doi              = {10.1103/physrevx.14.031038},
  modificationdate = {2026-08-06T18:37:32},
  publisher        = {American Physical Society (APS)},
}

@Article{Jiang2024,
  author           = {Jiang, Yi and Song, Ziyin and Zhu, Tiannian and Fang, Zhong and Weng, Hongming and Liu, Zheng-Xin and Yang, Jian and Fang, Chen},
  journal          = {Physical Review X},
  title            = {{Enumeration of Spin-Space Groups: Toward a Complete Description of Symmetries of Magnetic Orders}},
  year             = {2024},
  issn             = {2160-3308},
  month            = aug,
  number           = {3},
  pages            = {031039},
  volume           = {14},
  creationdate     = {2026-08-06T18:37:33},
  doi              = {10.1103/physrevx.14.031039},
  modificationdate = {2026-08-06T18:37:33},
  publisher        = {American Physical Society (APS)},
}

@Article{Xiao2024,
  author           = {Xiao, Zhenyu and Zhao, Jianzhou and Li, Yanqi and Shindou, Ryuichi and Song, Zhi-Da},
  journal          = {Physical Review X},
  title            = {{Spin Space Groups: Full Classification and Applications}},
  year             = {2024},
  issn             = {2160-3308},
  month            = aug,
  number           = {3},
  pages            = {031037},
  volume           = {14},
  creationdate     = {2026-08-06T18:37:33},
  doi              = {10.1103/physrevx.14.031037},
  modificationdate = {2026-08-06T18:37:33},
  publisher        = {American Physical Society (APS)},
}

@Article{Watanabe2024,
  author           = {Watanabe, Hikaru and Shinohara, Kohei and Nomoto, Takuya and Togo, Atsushi and Arita, Ryotaro},
  journal          = {Physical Review B},
  title            = {{Symmetry analysis with spin crystallographic groups: Disentangling effects free of spin-orbit coupling in emergent electromagnetism}},
  year             = {2024},
  issn             = {2469-9969},
  month            = mar,
  number           = {9},
  pages            = {094438},
  volume           = {109},
  creationdate     = {2026-08-06T18:37:33},
  doi              = {10.1103/physrevb.109.094438},
  modificationdate = {2026-08-06T18:37:33},
  publisher        = {American Physical Society (APS)},
}

@Article{Etxebarria2025,
  author           = {Etxebarria, Jesus and Perez-Mato, J. Manuel and Tasci, Emre S. and Elcoro, Luis},
  journal          = {Acta Crystallographica Section A Foundations and Advances},
  title            = {{Crystal tensor properties of magnetic materials with and without spin–orbit coupling. Application of spin point groups as approximate symmetries}},
  year             = {2025},
  issn             = {2053-2733},
  month            = jun,
  number           = {4},
  pages            = {317--338},
  volume           = {81},
  creationdate     = {2026-08-06T18:37:33},
  doi              = {10.1107/s2053273325004127},
  modificationdate = {2026-08-06T18:37:33},
  publisher        = {International Union of Crystallography (IUCr)},
}

@Article{Takenaka2005,
  author           = {Takenaka, K. and Takagi, H.},
  journal          = {Applied Physics Letters},
  title            = {{Giant negative thermal expansion in Ge-doped anti-perovskite manganese nitrides}},
  year             = {2005},
  issn             = {1077-3118},
  month            = dec,
  number           = {26},
  volume           = {87},
  creationdate     = {2026-08-06T18:37:33},
  doi              = {10.1063/1.2147726},
  modificationdate = {2026-08-06T18:47:22},
  publisher        = {AIP Publishing},
}

@Article{Takenaka2014,
  author           = {Takenaka, Koshi and Ichigo, Masayoshi and Hamada, Taisuke and Ozawa, Atsushi and Shibayama, Takashi and Inagaki, Tetsuya and Asano, Kazuko},
  journal          = {Science and Technology of Advanced Materials},
  title            = {{Magnetovolume effects in manganese nitrides with antiperovskite structure}},
  year             = {2014},
  issn             = {1878-5514},
  month            = feb,
  number           = {1},
  pages            = {015009},
  volume           = {15},
  creationdate     = {2026-08-06T18:37:33},
  doi              = {10.1088/1468-6996/15/1/015009},
  modificationdate = {2026-08-06T18:47:22},
  publisher        = {Informa UK Limited},
}

\end{document}